\documentclass[letterpaper, 10 pt, conference]{ieeeconf}  
\usepackage{color}
\usepackage{lineno}
\usepackage[cmex10]{amsmath}
\usepackage{amscd,amsfonts,amsmath,amssymb,mathrsfs,pifont,stmaryrd,tipa}
\usepackage{array,cases,dsfont,graphicx,texdraw}
\usepackage{graphics} 
\usepackage{epsfig} 
\usepackage{mathbbold}
\usepackage{stfloats}
\usepackage{subfig}
\usepackage{hyperref}
\IEEEoverridecommandlockouts
\newtheorem{Remark}{Remark}
\newtheorem{Corollary}{Corollary}
\newtheorem{Definition}{Definition}
\newtheorem{Problem}{Problem}
\newenvironment{Proof}{\noindent{\em Proof:\/}}{\hfill $\Box$\par}
\newtheorem{Theorem}{Theorem}
\newtheorem{Lemma}{Lemma}

\newtheorem{Assumption}{Assumption}

\input{tcilatex}                                                          

\IEEEoverridecommandlockouts                              
\title{\LARGE \bf
Game Design and Analysis for Price based Demand Response: An Aggregate Game Approach}

\author{Maojiao Ye, \emph{Student Member, IEEE,}
        and Guoqiang Hu, \emph{Member, IEEE}
\thanks{M. Ye and G. Hu are with the School of Electrical and Electronic Engineering, Nanyang
Technological University, 639798, Singapore  (Email: mjye@ntu.edu.sg, gqhu@ntu.edu.sg).}
\thanks{This work was supported by Singapore Economic Development Board under EIRP grant S14-1172-NRF EIRP-IHL.}}

\begin{document}

\maketitle
\thispagestyle{empty}
\pagestyle{empty}

\begin{abstract}

In this paper, an aggregate game is adopted for the modeling and analysis of energy consumption control in smart grid. Since the electricity users' cost functions depend on the aggregate energy consumption, which is unknown to the end users, an average consensus protocol is employed to estimate it. By neighboring communication among the users about their estimations on the aggregate energy consumption, Nash seeking strategies are developed. Convergence properties are explored for the proposed Nash seeking strategies. For energy consumption game that may have multiple isolated Nash equilibria, a local convergence result is derived. The convergence is established by utilizing singular perturbation analysis and Lyapunov stability analysis. Energy consumption control for a network of heating, ventilation, and air conditioning (HVAC) systems is investigated. Based on the uniqueness of the Nash equilibrium, it is shown that the players' actions converge to a neighborhood of the unique Nash equilibrium non-locally. More specially, if the unique Nash equilibrium is an inner Nash equilibrium, an exponential convergence result is obtained. Energy consumption game with stubborn players is studied. In this case, the actions of the rational players can be driven to a neighborhood of their best response strategies by using the proposed method. Numerical examples are presented to verify the effectiveness of the proposed methods.
\end{abstract}


\section{INTRODUCTION}
Demand response schemes are designed to motivate the electricity users to adjust their energy consumptions to the desired profile based on supply conditions \cite{Gelazanskas14}. Due to the benefits, such as improving system reliability, efficiency, security and customer bill savings, great efforts have been dedicated to control and optimization problems related to demand response in smart grid (e.g., see \cite{Vardakas14}-\cite{YETCST} and the references therein). The electricity users are categorized as price-takers and price-anticipating users in the existing literatures \cite{PEDRAM12}. The price-takers schedule their energy consumptions regardless of their effect on the electricity price. In contrast, the price-anticipating users consider the impact of their energy consumptions on the electricity price. Industrial users and commercial users with large energy consumption are typical examples of price-anticipating users. Taking the price as a function of the aggregate energy consumption brings up coupled energy consumption control problems among the electricity users.

Game theory is an effective modeling and analysis tool to deal with the interactions among the electricity consumers \cite{Vardakas14}.
Game theoretical approaches such as stackelberg game, evolutionary game, differential game, just to name a few, have been extensively utilized to design and analyze demand response schemes in smart grid (e.g., \cite{Nekouer14}-\cite{Maharjan13}).
For example, in \cite{Nekouer14}, Stackelberg game was leveraged to model the interaction between the demand response aggregators and generators. The Nash equilibrium of the game among the generators was computed via solving a centralized quadratic optimization problem. In \cite{Amir2010}, the authors considered peak-to-average ratio (PAR) minimization and energy cost minimization by designing a non-cooperative energy consumption game. By communicating among the players on their scheduled daily energy consumptions, a Nash seeking strategy was proposed. A coupled-constraint game was used in \cite{Deng14} for scheduling residential energy consumption. Noticing that the best response algorithm suffers from privacy concern, a gradient projection method based on estimations of the price changing trend was proposed at the cost of computation time and memory. Multiple utility companies were considered in \cite{Chai14} where the authors used a two-level game approach to manage the interactions. The competition among the utility companies was described by a non-cooperative game. The users were modeled as evolutionary game players. Considering the system dynamics, a differential game approach was proposed in \cite{forouzandehmehr14}. In this paper, the energy consumption control is considered as an aggregate game
\cite{Jensen10}-\cite{Martimort} on graph.

\textbf{Related Work:} The main objective of this paper is to achieve Nash equilibrium seeking (e.g., see \cite{ZhuCDC12}-\cite{Salehisadaghiani14} and the references therein for an incomplete reference list) in energy consumption game. Different from many Nash seeking strategies where full communication among the players is used, in \cite{Koshal12} and \cite{Salehisadaghiani14}, the Nash equilibrium is attained by utilizing neighboring communication.  An aggregate game (e.g., see \cite{Jensen10}-\cite{Koshal12}) in which the players interact through the sum of the players' actions was investigated in \cite{Koshal12}.  By utilizing a gossip algorithm, the players can search for the Nash equilibrium through neighboring communication. This idea was further generalized in \cite{Salehisadaghiani14}. The players were considered to be generally interacting with each other. The game on graph was then solved via using a gossip algorithm. Similar to \cite{Koshal12} and \cite{Salehisadaghiani14}, this paper leverages the idea of game on graph to solve energy consumption game in smart grid.

This paper aims to solve energy consumption control for a network of price-anticipating electricity users. Compared with the existing works, the main contributions of the paper are summarized as follows.
\begin{itemize}
  \item An aggregate game is adopted for the modeling and analysis of energy consumption control in smart grid. By using game on graph, the users update their actions based on the communication with their neighbors. This scheme reduces the communication between the electricity users with the centralized agent (e.g., energy provider). Hence, the single-node congestion problem is relieved.
  \item An aggregate game that may admit multiple isolated Nash equilibria is firstly considered. Based on an average consensus protocol, a Nash seeking strategy is proposed for the players. An energy consumption game for a network of heating, ventilation and air conditioning (HVAC) systems, in which the Nash equilibrium is unique, is investigated. The Nash seeking strategy is designed based on a primal-dual dynamics. More specifically, if the unique Nash equilibrium is an inner Nash equilibrium, an exponential convergence result is derived. Energy consumption game with stubborn players is studied. It is shown that with the presence of stubborn players, the rational players' actions converge to a neighborhood of their best response strategies. The proposed Nash seeking strategies serve as alternative approaches for the gossip-algorithm in \cite{Koshal12} to solve aggregative game on graph.
  \item The end-users only need to communicate with their neighbors on their estimations of the aggregate energy consumption. They don't need to share their own energy consumptions with their opponents. Hence, the privacy of the electricity users is protected.
\end{itemize}

The rest of the paper is structured as follows. In Section \ref{pre_lim}, some related preliminaries are provided. System model and the problem formulation are given in Section \ref{sys_pro}. A general energy consumption game that may have multiple Nash equilibria is studied in Section \ref{main_res_lcc} without considering the constraints. Energy consumption game among a network of HVAC systems is investigated in Section \ref{inner_na_se}. In Section \ref{num_e}, numerical examples are provided to verify the effectiveness of the proposed methods. Conclusions and future directions are stated in Section \ref{con_a}.

\section{Preliminaries and Notations}\label{pre_lim}

In this paper, $R$ represents for the set of real numbers, $R_+$ stands for the set of non-negative real numbers and $R_{++}$ is the set of positive real numbers. Furthermore, diag$\{k_i\}$ for $i\in\{1,2,\cdots,N\}$ is defined as
\begin{equation}
\text{diag}\{k_i\}=\left[
                                                                                                                 \begin{array}{cccc}
                                                                                                                   k_1 & 0 & \cdots & 0 \\
                                                                                                                   0 & k_2 &  & \\
                                                                                                                   \vdots &  & \ddots & \vdots \\
                                                                                                                   0 & \cdots &  & k_N \\
                                                                                                                 \end{array}
                                                                                                               \right],\nonumber
\end{equation}
and $\left[h_i\right]_{vec}$ for $i\in \{1,2,\cdots,N\}$ is defined as $\left[h_i\right]_{vec}=[h_1,h_2,\cdots,h_N]^T.$

\subsection{Game Theory}
Below are some definitions on game theory \cite{Martimort}-\cite{Monderer96}. 
\begin{Definition}
A game in normal form is defined as a triple
$\Gamma\triangleq\{\mathbb{N},X,C\}$ where $\mathbb{N}=\{1,2,\cdots,N\}$ is the set of $N$ players, $X=X_1\times\cdots \times X_N$, $X_i\subset R$ is the set of actions for
player $i$, and $C=(C_{1},\cdots,C_{N})$ where $C_{i}$ is the cost function of player $i$.
\end{Definition}

\begin{Definition}
\emph{(Potential game)} A game $\Gamma$ is a potential game if there exists a function $P$ such that
\begin{equation}\label{pote_ga_de}
\frac{\partial C_i(l_i,\mathbf{l}_{-i})}{\partial l_i}=\frac{\partial P(l_i,\mathbf{l}_{-i})}{\partial l_i},
\end{equation}
$\forall i \in \mathds{N}.$ Furthermore, the function $P$ is the potential function. In \eqref{pote_ga_de}, $l_i$ denotes the action of player $i$ and, $\mathbf{l}_{-i}$ denotes all the players' actions
other than the action of player $i,$ i.e., $\mathbf{l}_{-i}=[l_1,\cdots,l_{i-1},l_{i+1},\cdots,l_N]^T.$
\end{Definition}
\begin{Definition}
\emph{(Aggregate game)}
An aggregate game is a normal form game with the players' cost functions depending only on its own action and a linear aggregate of the full action profile.
\end{Definition}
\begin{Definition}
\emph{(Nash equilibrium)} Nash equilibrium is an action profile
on which no player can reduce its cost by unilaterally changing its own
action, i.e., an action profile $\mathbf{l}^*=(l_{i}^{\ast},\mathbf{l}_{-i}^{\ast})\in X$ is the
Nash equilibrium if for all the players
\begin{equation}
C_{i}(l_{i}^{\ast},\mathbf{l}%
_{-i}^{\ast})\leq C_{i}(l_{i},\mathbf{l}_{-i}^{\ast}),\forall i\in \mathds{N}
\end{equation}
 for all $l_i\in X_i$.

For an aggregate game $\Gamma$ with the aggregate function being $\bar{l}=\sum_{i=1}^{N}l_i$, an equilibrium of the aggregate game $\Gamma$ is an action profile $\mathbf{l}^*\in X$ on which $\forall i\in \mathds{N}$
\begin{equation*}
C_i(l_i^*,l_i^*+\sum_{j=1,j\neq i}^N l_j^*)\leq C_i(l_i,l_i+\sum_{j=1,j\neq i}^N l_j^*),\forall l_i\in X_i
\end{equation*}
and the associated equilibrium aggregate is denoted as $\bar{l}^*=\sum_{i=1}^{N}l_i^*.$
\end{Definition}

\subsection{Graph Theory }

For a graph defined as $G=(V,E)$ where $E$ is the edge set satisfying
$E\subset V\times V$ with $V=\{1,2,\cdots,N\}$ being the set of nodes in the
network, it is undirected if for every $(i,j)\in E,$ $(j,i)\in
E.$ An undirected graph is connected if there exists a path between any pair
of distinct vertices. The element in the adjacency matrix $A$ is defined as
$a_{ij}=1$ if node $j$ is connected with node $i,$ else, $a_{ij}=0.$ The neighboring set of agent $i$ is defined as $\mathcal{N}_i=\{j \in V|(j,i)\in E\}.$
The Laplacian matrix for the graph $L$ is defined as $L=\bar{D}-A $ where $\bar{D}$ is
defined as a diagonal matrix whose $i$th diagonal element is equal to the out
degree of node $i,$ represented by $\sum_{j=1}^{N}a_{ij}$.

\subsection{Dynamic Average Consensus}
\begin{Theorem}\label{The_1}
\cite{Freeman06,Menon14} Let $G$ be a connected, undirected
graph, $L$ be the Laplacian of $G$. Then, for any constant $\mathbf{\omega}\in R^{N}$, the state of the following system
\[
\left[
\begin{array}
[c]{c}%
\dot{\mathbf{x}}\\
\dot{\mathbf{y}}%
\end{array}
\right]  =\left[
\begin{array}
[c]{cc}%
-I-L & -L\\
L & \mathbf{0}
\end{array}
\right]  \left[
\begin{array}
[c]{c}%
\mathbf{x}\\
\mathbf{y}
\end{array}
\right]  +\left[
\begin{array}
[c]{c}%
\mathbf{\omega}\\
\mathbf{0}
\end{array}
\right],
\]
with arbitrary initial conditions $\mathbf{x}(0),\mathbf{y}(0)\in R^{N}$ remains bounded
and $\mathbf{x}(t)$ converges exponentially to $\frac{1}{N}\mathbf{1}^{T}\mathbf{\omega}
\mathbf{1}$ as $t\rightarrow\infty$ where $\mathbf{1}$ denotes an $N$
dimensional column vector composed of $1$.
\end{Theorem}
\subsection{Saddle Point}
The pair $(\mathbf{x}^*,\mathbf{y}^*)$ is the saddle point of $F(\mathbf{x},\mathbf{y})$ if \begin{equation}
F(\mathbf{x}^*,\mathbf{y})\leq F(\mathbf{x}^*,\mathbf{y}^*) \leq F(\mathbf{x},\mathbf{y}^*),
\end{equation}
is satisfied \cite{YETCST,YECDC15}.
\section{System Model and Problem Formulation}\label{sys_pro}

\begin{figure}
\begin{center}
\scalebox{0.62}{\includegraphics[65,341][379,641]{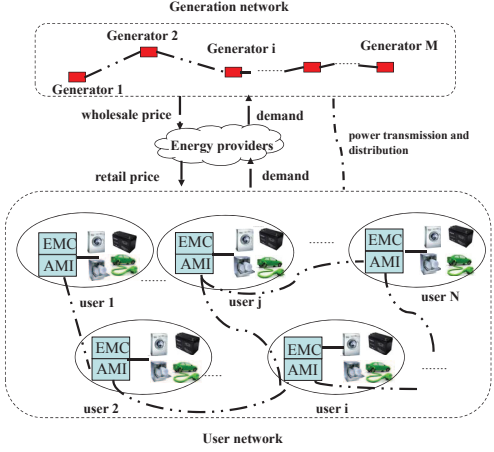}}
\caption{A simplified illustration of the electricity buying and selling model.}\label{Fig1}
\end{center}
\end{figure}

Consider an electricity market with a network of $N$ price-anticipating users. The simplified illustration of the electricity buying and selling model is shown in Fig. \ref{Fig1}.

The users are equipped with an energy-management controller (EMC) and an
advanced metering infrastructure (AMI) \cite{MaHuSpanos}. The EMC is used to schedule the electricity usage
for the corresponding user. The AMI enables bidirectional
communication among the electricity users and the centralized agent (e.g., the energy provider). The communication between the electricity users and their neighbors can be modeled by an undirected and connected graph. In this paper, we suppose that the electricity users schedule their energy consumption by minimizing their own costs.

Let $l_{i}$ be the energy consumption of user $i.$ Then, the cost of user $i$ can be quantified as \cite{MaHuSpanos},
\begin{equation}
C_{i}(l_{i},\bar{l})=V_{i}(l_{i})+P(\bar{l})l_{i},\label{DR_cost_model}%
\end{equation}
where $\bar{l}$ is the aggregate energy consumption of all the electricity users, i.e.,
$\bar{l}=%
{\textstyle\sum_{j=1}^{N}}
l_{j}$. Furthermore, $V_{i}(l_{i})$ is the load curtailment cost \cite{MaHuSpanos}. The term $P(\bar{l})l_{i}$ represents the billing
payment for the consumption of energy $l_{i}$ where the price
$P(\bar{l})$ is a function of the aggregate energy consumption $\bar{l}.$

For each user, the energy consumption should be within its acceptable range,
i.e., $l_{i}\in\lbrack l_{i}^{\min},l_{i}^{\max}]$ where $l_{i}^{\min}<l_{i}^{\max}$ and $l_{i}^{\min},$
$l_{i}^{\max}$ are the minimal and maximal acceptable energy consumption for
user $i,$ respectively.
The problem is defined as follows.
\begin{Problem}
\label{DR_CO}\emph{(Strategy Design for the Electricity Consumers in the Energy Consumption Game)}
In the energy consumption game, the electricity users are the players. The energy consumption $l_i$ and $C_i(l_i,\bar{l})$ are the action and cost of player $i$, respectively. Player $i$'s objective is defined as
\begin{align*}
& \min_{l_{i}}\text{ }C_{i}(l_{i},\bar{l})\label{DR_cent_opt}\\
& \text{subject to }l_{i}^{\min}\leq l_{i}\leq l_{i}^{\max}, i\in \mathds{N},
\end{align*}
where $\mathds{N}=\{1,2,\cdots,N\}$ denotes the set of electricity users. The aggregate energy consumption $\bar{l}$ is unknown to the players. Suppose that the pure-strategy Nash equilibrium exists and is isolated. Furthermore, $C_i, i\in \mathds{N}$ are smooth functions over $R^N$. Design a control strategy for the players such that they can search for the Nash equilibrium.
\end{Problem}
\begin{Remark}
In practice, providing the aggregate energy consumption $\bar{l}$ to all the users is challenging for the centralized agent when the users are dynamically updating their actions. Hence, we consider the Nash equilibrium seeking under the condition that the
users have no access to the aggregate energy consumption $\bar{l}$. But the users are allowed to communicate with
their neighbors on the estimations of the aggregate energy consumption. Furthermore, we suppose that the centralized agent can broadcast the total number of electricity users $N$ to all the electricity users in the network.
\end{Remark}

In summary, the energy consumption control considered in this paper is based on the following assumption.
\begin{Assumption}\label{Ass_model}
The electricity users can communicate with their neighbors via an undirected and connected graph. Furthermore, the total number of the electricity users, $N$, is known to all the electricity users.
\end{Assumption}
\section{Energy Consumption Game Design and Analysis}\label{main_res_lcc}

In this section, the energy consumption game is considered in the general form (the pricing function is not specified). For simplicity, the constraints $l_{i}^{\min}\leq l_{i}\leq l_{i}^{\max}, i\in \mathds{N}$, are not considered in this section.
\subsection{Game Analysis}\label{Sec_GA}

Before we proceed to facilitate the subsequent analysis, the following assumptions are made.

\begin{Assumption}\label{Assum_1}
\cite{Frihauf12} There exists isolated, stable Nash equilibrium on which
\begin{equation}
\begin{aligned}
&\frac{\partial C_i}{\partial l_i}(\mathbf{l}^*)=0,\\
&\frac{\partial ^2C_i}{\partial l_i^2}(\mathbf{l}^*)>0, \forall i\in \mathds{N}
\end{aligned}
\end{equation}
where $\mathbf{l}^*$ denotes the Nash equilibrium.
\end{Assumption}

\begin{Assumption}\label{Assum_2}
\cite{Frihauf12} The matrix
\begin{equation}
B=\left[
                \begin{array}{cccc}
                  \frac{\partial ^2C_1}{\partial l_1^2}(\mathbf{l}^*) & \frac{\partial ^2C_1}{\partial l_1 \partial l_2}(\mathbf{l}^*) & \cdots &  \frac{\partial ^2C_1}{\partial l_1 \partial l_N}(\mathbf{l}^*) \\
                  \frac{\partial ^2C_2}{\partial l_1 \partial l_2}(\mathbf{l}^*) & \frac{\partial ^2C_2}{\partial l_2^2}(\mathbf{l}^*) &  & \vdots \\
                  \vdots &  & \ddots &  \\
                  \frac{\partial ^2C_N}{\partial l_1 \partial l_N}(\mathbf{l}^*) & \cdots &  & \frac{\partial ^2C_N}{\partial l_N^2}(\mathbf{l}^*) \\
                \end{array}
              \right]
\end{equation}
is strictly diagonally dominant.
\end{Assumption}

Inspired by \cite{Koshal12}, let $D_i$ denote player $i$'s estimation on the aggregate energy consumption. By using the estimations, the players' objectives can be rewritten as:

\begin{Problem}\label{prob3}Player $i$'s objective is
\begin{equation}
\min_{l_{i}}\text{ }C_{i}(l_{i},D_i),
\end{equation}
where $D_i=\sum_{j=1}^{N}l_{j}, \forall i\in \mathds{N}.$
\end{Problem}

In the following, a consensus based method will be proposed to search for the Nash equilibrium of the energy consumption game (without considering the constraints).

\subsection{Nash Equilibrium Seeking for the Aggregate Energy Consumption Game}\label{Nas_ex}

Based on the consensus protocol in \cite{Freeman06}-\cite{Menon14}, the Nash seeking strategy for player $i,i\in \mathds{N}$ is designed as

\begin{subequations}\label{upd_law}
\begin{align}
&\dot{D}_i=-D_i-\sum_{j \in \mathcal{N}_i}(D_{i}-D_{j})-\sum_{j \in \mathcal{N}_i}(\kappa_{i}-\kappa_{j})+Nl_i\label{upd_law_1}\\
&\dot{\kappa}_{i}=\sum_{j \in \mathcal{N}_i}(D_{i}-D_{j})\label{upd_law_3}\\
&\dot{l}_i=-\bar{k}_i(\frac{\partial V_i}{\partial l_i}+P(D_i)+l_i\frac{\partial P(D_i)}{\partial D_i}),
\end{align}
\end{subequations}
where $\mathcal{N}_i$ denotes the neighboring set of player $i$, $\bar{k}_i=\delta k_i, i\in \mathds{N}$, $\delta$ is a small positive parameter and $k_i$ is a fixed positive parameter. Furthermore, $\kappa_{i},i\in \mathds{N}$ are intermediate variables.

Writing \eqref{upd_law} in the concatenated form gives
\begin{subequations}\label{upd_law_c}
\begin{align}
    \left[
    \begin{array}{c}
    \dot{\mathbf{D}} \\
    \dot{\mathbf{\kappa}}
    \end{array}
    \right]=&\left[
    \begin{array}{cc}
    -I-L & -L \\
    L & \mathbf{0} \\
    \end{array}
    \right]
    \left[
    \begin{array}{c}
    \mathbf{D} \\
    \mathbf{\kappa} \\
    \end{array}
    \right]+\left[
    \begin{array}{c}
    N\mathbf{l} \\
    \mathbf{0} \\
    \end{array}
    \right]\label{upd_law_c_1} \\
    \dot{\mathbf{l}}=&-\delta \mathbf{\mathbf{k}} ([\frac{\partial V_i}{\partial l_i}+P(D_i)+l_i\frac{\partial P(D_i)}{\partial D_i}]_{vec}),\label{upd_law_c_2}
    \end{align}
\end{subequations}
where $\mathbf{k}$ is defined as $\mathbf{k}=\text{diag}\{k_i\}$, $i\in \mathds{N}$, $\mathbf{D}$, $\mathbf{\kappa}$, $\mathbf{l}$ are the concatenated vectors of $D_i$, $\kappa_i$, $l_i$, respectively. Let $U=[\widetilde{U}, \mu]$ be an $N\times N$ dimensional orthonormal matrix such that $\mu^T L=0$ where $\mu$ is an $N$ dimensional column vector. Furthermore, let $\mathbf{\kappa}=U \left[ \begin{array}{c} \widetilde{\mathbf{\kappa}} \\ \kappa_\mu \end{array} \right] $  where $\widetilde{\kappa}$ is an $N-1$ dimensional column vector \cite{YETCST}. Then, the closed-loop system can be rewritten as
\begin{subequations}\label{upd_law_ch}
\begin{align}
    \left[
    \begin{array}{c}
    \dot{\mathbf{D}} \\
    \dot{\widetilde{\mathbf{\kappa}}}
    \end{array}
    \right]&=\left[
    \begin{array}{cc}
    -I-L & -L \widetilde{U} \\
    \widetilde{U}^T L & \mathbf{0} \\
    \end{array}
    \right]
    \left[
    \begin{array}{c}
    \mathbf{D} \\
    \widetilde{\mathbf{\kappa}} \\
    \end{array}
    \right]+\left[
    \begin{array}{c}
    N\mathbf{l} \\
    \mathbf{0} \\
    \end{array}
    \right]\label{upd_law_ch_1} \\
    \dot{\mathbf{l}}&=-\delta \mathbf{\mathbf{k}} ([\frac{\partial V_i}{\partial l_i}+P(D_i)+l_i\frac{\partial P(D_i)}{\partial D_i}]_{vec}),\label{upd_law_ch_3}
    \end{align}
\end{subequations}
and
\begin{equation}
\dot{\kappa}_\mu=0,
\end{equation}
which indicates that $\kappa_\mu(t)=\kappa(0).$

Suppose that $\mathbf{D}^e(\mathbf{l})$,  $\widetilde{\mathbf{\kappa}}^e(\mathbf{l})$ are the quasi-steady states of $\mathbf{D}$ and $\widetilde{\mathbf{\kappa}}$, respectively, i.e.,
\begin{equation}
\left[
    \begin{array}{cc}
    -I-L & -L \widetilde{U} \\
    \widetilde{U}^T L & \mathbf{0} \\
    \end{array}
    \right]
    \left[
    \begin{array}{c}
    \mathbf{D}^e(\mathbf{l}) \\
    \widetilde{\mathbf{\kappa}}^e(\mathbf{l}) \\
    \end{array}
    \right]+\left[
    \begin{array}{c}
    N\mathbf{l} \\
    \mathbf{0} \\
    \end{array}
    \right]=\mathbf{0},
\end{equation}
for fixed $\mathbf{l}$.
Note that $\mathbf{D}^e(\mathbf{l})$,  $\widetilde{\mathbf{\kappa}}^e(\mathbf{l})$ is unique for fixed $\mathbf{l}$ as the matrix $\left[
    \begin{array}{cc}
    -I-L & -L \widetilde{U} \\
    \widetilde{U}^T L & \mathbf{0} \\
    \end{array}
    \right]$ is Hurwitz \cite{Menon14}.
Then, by direct calculation, it can be derived that

\begin{equation}
\left[
    \begin{array}{cc}
    -I-L & -L \\
    L & \mathbf{0} \\
    \end{array}
    \right]
    \left[
    \begin{array}{c}
    \mathbf{D}^e(\mathbf{l}) \\
    U[(\widetilde{\mathbf{\kappa}}^{e}(\mathbf{l}))^T,\kappa_\mu]^T \\
    \end{array}
    \right]+\left[
    \begin{array}{c}
    N\mathbf{l} \\
    \mathbf{0} \\
    \end{array}
    \right]=\mathbf{0},
\end{equation}
which indicates that $\mathbf{D}^e(\mathbf{l})$ is the equilibrium of the original system. Hence, by using Theorem \ref{The_1}, it can be concluded that $\mathbf{D}^e(\mathbf{l})=\mathbf{1}\sum_{i=1}^{N}l_i$ for fixed $\mathbf{l}$.

\begin{Theorem}\label{res_1}
Suppose that Assumptions \ref{Ass_model}-\ref{Assum_2} hold. Then, there exists a positive constant $\delta^*$ such that for every $0<\delta<\delta^*$,
$(\mathbf{l}(t),\mathbf{D}(t),\widetilde{\mathbf{\kappa}}(t))$ converges exponentially to  $(\mathbf{l}^*,\mathbf{1}\sum_{i=1}^{N}l_i^*,\widetilde{\mathbf{\kappa}}^e(\mathbf{l}^*))$
under \eqref{upd_law_ch} given that $||\mathbf{l}(0)-\mathbf{l}^*||$, $||\mathbf{D}(0)-\mathbf{1}\sum_{i=1}^{N}l_i^*||$, $||\widetilde{\mathbf{\kappa}}(0)-\widetilde{\kappa}^e(\mathbf{l}^*)||$ are sufficiently small.
\end{Theorem}

\begin{Proof}
Let $\tau=\delta t$. Then, in the $\tau$-time scale, the reduced-system is
\begin{subequations}\label{upd_law_chhh}
\begin{align}
    \left[
    \begin{array}{c}
    \delta \frac{d\mathbf{D}}{d\tau} \\
    \delta \frac{d\widetilde{\mathbf{\kappa}}}{d\tau}
    \end{array}
    \right]&=\left[
    \begin{array}{cc}
    -I-L & -L \widetilde{U} \\
    \widetilde{U}^T L & \mathbf{0} \\
    \end{array}
    \right]
    \left[
    \begin{array}{c}
    \mathbf{D} \\
    \widetilde{\mathbf{\kappa}} \\
    \end{array}
    \right]+\left[
    \begin{array}{c}
    N\mathbf{l} \\
    \mathbf{0} \\
    \end{array}
    \right]\label{upd_law_chhh_1} \\
    \frac{d\mathbf{l}}{d\tau}&=-\mathbf{k} ([\frac{\partial V_i}{\partial l_i}+P(D_i)+l_i\frac{\partial P(D_i)}{\partial D_i}]_{vec}).\label{upd_law_chhh_3}
    \end{align}
\end{subequations}

\emph{Quasi-steady state analysis}: letting $\delta=0$ freezes $\mathbf{D}$ and $\widetilde{\mathbf{\kappa}}$ at the quasi-steady state on which $D_i=\sum_{j=1}^{N} l_j,\forall i\in \mathds{N}$. Hence, the reduced-system is
\begin{equation}\label{red_sy_2}
\begin{aligned}
\frac{d\mathbf{l}}{d\tau}=&- \mathbf{\mathbf{k}} ([\frac{\partial V_i}{\partial l_i}+P(D_i)+l_i\frac{\partial P(D_i)}{\partial D_i}]_{vec})\\
    =&-\mathbf{k} ([\frac{\partial C_i(l_i,\mathbf{l}_{-i})}{\partial l_i}]_{vec}).
\end{aligned}
\end{equation}
Linearizing \eqref{red_sy_2} at the Nash equilibrium point $\mathbf{l}^*$ gives,
\begin{equation}\label{red_sy_1}
\frac{d\mathbf{l}}{d\tau}=-\mathbf{k}B(\mathbf{l}-\mathbf{l}^*),
\end{equation}
where $-\mathbf{k}B$ is Hurwitz by Assumption \ref{Assum_2} and the Gershgorin Circle Theorem \cite{Horn85}.
Hence, the equilibrium point $\mathbf{l}^*$ is locally exponentially stable under \eqref{red_sy_2}, i.e., there exist positive constants $\varrho_1$ and $\varrho_2$ such that (denote the trajectory of \eqref{red_sy_2} as $\mathbf{l}_r(\tau)$)
\begin{equation}
||\mathbf{l}_r(\tau)-\mathbf{l}^*||\leq \varrho_1 e^{-\varrho_2 \tau}||\mathbf{l}_r(0)-\mathbf{l}^*||,
\end{equation}
given that $||\mathbf{l}_r(0)-\mathbf{l}^*||$ is sufficiently small.

\emph{Boundary-layer analysis}: Since $(\mathbf{D}^e(\mathbf{l}),\widetilde{\mathbf{\kappa}}^e(\mathbf{l}))$ satisfies,
\begin{equation}
\left[
    \begin{array}{cc}
    -I-L & -L \widetilde{U} \\
    \widetilde{U}^T L & \mathbf{0} \\
    \end{array}
    \right]
    \left[
    \begin{array}{c}
    \mathbf{D}^e(\mathbf{l}) \\
    \widetilde{\mathbf{\kappa}}^e(\mathbf{l}) \\
    \end{array}
    \right]+\left[
    \begin{array}{c}
    N\mathbf{l} \\
    \mathbf{0} \\
    \end{array}
    \right]=\mathbf{0},
\end{equation}
it can be derived that $\mathbf{D}^e(\mathbf{l})$, $\widetilde{\mathbf{\kappa}}^e(\mathbf{l})$ are linear functions of $\mathbf{l}$ as the matrix $\left[
    \begin{array}{cc}
    -I-L & -L \widetilde{U} \\
    \widetilde{U}^T L & \mathbf{0} \\
    \end{array}
    \right]$ is non-singular.

Let
\begin{equation}
\begin{aligned}
&\mathbf{D}^{'}=\mathbf{D}-\mathbf{D}^e\\
&\widetilde{\mathbf{\kappa}}^{'}=\mathbf{\kappa}-\mathbf{\kappa}^e.
\end{aligned}
\end{equation}

Then,
\begin{equation}
\begin{aligned}
\delta\left[
    \begin{array}{c}
    \frac{d\mathbf{D}^{'}}{d\tau} \\
    \frac{d\widetilde{\mathbf{\kappa}}^{'}}{d\tau}
    \end{array}
    \right]&=\left[
    \begin{array}{cc}
    -I-L & -L \widetilde{U} \\
    \widetilde{U}^T L & \mathbf{0} \\
    \end{array}
    \right]
    \left[
    \begin{array}{c}
    \mathbf{D}'+\mathbf{D}^e \\
    \widetilde{\mathbf{\kappa}}'+\widetilde{\mathbf{\kappa}}^e \\
    \end{array}
    \right]\\
    &+\left[
    \begin{array}{c}
    N\mathbf{l} \\
    \mathbf{0} \\
    \end{array}
    \right]-\delta\left[
    \begin{array}{c}
    \frac{\partial \mathbf{D}^{eT}}{\partial \mathbf{l}}^T \\
    \frac{\partial \widetilde{\mathbf{\kappa}}^{eT}}{\partial \mathbf{l}}^T
    \end{array}
    \right]\\
    &\times(-\mathbf{\mathbf{k}} ([\frac{\partial V_i}{\partial l_i}+P(D_i'+\sum_{j=1}^{N}l_j)\\
    &+l_i\frac{\partial P(D_i'+\sum_{j=1}^{N}l_j)}{\partial (D_i'+\sum_{j=1}^{N}l_j)}]_{vec}))\\
    &=\left[
    \begin{array}{cc}
    -I-L & -L \widetilde{U} \\
    \widetilde{U}^T L & \mathbf{0} \\
    \end{array}
    \right]
    \left[
    \begin{array}{c}
    \mathbf{D}' \\
    \widetilde{\mathbf{\kappa}}' \\
    \end{array}
    \right]\\
    &-\delta\left[
    \begin{array}{c}
    \frac{\partial \mathbf{D}^{eT}}{\partial \mathbf{l}}^T \\
    \frac{\partial \widetilde{\mathbf{\kappa}}^{eT}}{\partial \mathbf{l}}^T
    \end{array}
    \right](-\mathbf{\mathbf{k}} ([\frac{\partial V_i}{\partial l_i}+P(D_i'+\sum_{j=1}^{N}l_j)\\
    &+l_i\frac{\partial P(D_i'+\sum_{j=1}^{N}l_j)}{\partial (D_i'+\sum_{j=1}^{N}l_j)}]_{vec})).
\end{aligned}
\end{equation}

Hence, in $t$-time scale

\begin{equation}
\begin{aligned}
\left[
    \begin{array}{c}
    \frac{d\mathbf{D}^{'}}{dt} \\
    \frac{d\widetilde{\mathbf{\kappa}}^{'}}{dt}
    \end{array}
    \right]=&\left[
    \begin{array}{cc}
    -I-L & -L \widetilde{U} \\
    \widetilde{U}^T L & \mathbf{0} \\
    \end{array}
    \right]
    \left[
    \begin{array}{c}
    \mathbf{D}' \\
    \widetilde{\mathbf{\kappa}}' \\
    \end{array}
    \right]\\
    &-\delta\left[
    \begin{array}{c}
    \frac{\partial \mathbf{D}^{eT}}{\partial \mathbf{l}}^T \\
    \frac{\partial \widetilde{\mathbf{\kappa}}^{eT}}{\partial \mathbf{l}}^T
    \end{array}
    \right](-\mathbf{\mathbf{k}} ([\frac{\partial V_i}{\partial l_i}\\
    &+P(D_i'+\sum_{j=1}^{N}l_j)\\
    &+l_i\frac{\partial P(D_i'+\sum_{j=1}^{N}l_j)}{\partial (D_i'+\sum_{j=1}^{N}l_j)}]_{vec})).
\end{aligned}
\end{equation}

Letting $\delta=0$ gives the boundary-layer model of \eqref{upd_law_chhh} as

\begin{equation}
\left[
    \begin{array}{c}
    \frac{d\mathbf{D}^{'}}{dt} \\
    \frac{d\widetilde{\mathbf{\kappa}}^{'}}{dt}
    \end{array}
    \right]=\left[
    \begin{array}{cc}
    -I-L & -L \widetilde{U} \\
    \widetilde{U}^T L & \mathbf{0} \\
    \end{array}
    \right]
    \left[
    \begin{array}{c}
    \mathbf{D}^{'} \\
    \widetilde{\mathbf{\kappa}}^{'} \\
    \end{array}
    \right].
\end{equation}
Since the matrix $\left[
    \begin{array}{cc}
    -I-L & -L \widetilde{U} \\
    \widetilde{U}^T L & \mathbf{0} \\
    \end{array}
    \right]$ is Hurwitz, the equilibrium point of the boundary-layer model $\mathbf{D}^{'}=0,\widetilde{\mathbf{\kappa}}^{'}=0$ is exponentially stable, uniformly in $(t,\mathbf{l})$.

Therefore, by Theorem 11.4 in \cite{KHAIL02}, it can be concluded that there exists a positive constant $\delta^*$ such that for all $0 < \delta < \delta^*$, $(\mathbf{l}^*,\mathbf{1}\sum_{i=1}^{N}l_i^*,\widetilde{\mathbf{\kappa}}^e(\mathbf{l}^*))$ is exponentially stable under \eqref{upd_law_ch}.

\end{Proof}

From the analysis, it can be seen that all the states in \eqref{upd_law} stay bounded and $\mathbf{l}(t)$ produced by \eqref{upd_law} converges to the Nash equilibrium under the given conditions.

In this section, a Nash seeking strategy is proposed without requiring the uniqueness of the Nash equilibrium. In the following section, energy consumption game for HVAC systems where the Nash equilibrium is unique will be considered.

\section{Energy Consumption Game among A Network of HVAC Systems}\label{inner_na_se}

For HVAC systems, the load curtailment cost may be modeled as \cite{MaHuSpanos}
\begin{equation}
V_i(l_i)=\nu_i\xi_i^2(l_i-\hat{l}_i)^2,i\in \mathds{N},
\end{equation}
where $\nu_i$ and $\xi_i$ are thermal coefficients, $\nu_i\xi_i^2>0$ and $\hat{l}_i$ is the energy needed to maintain the indoor temperature of the HVAC system. When the pricing function is
\begin{equation}
P(\bar{l})=a\sum_{i=1}^N l_i+p_0,
\end{equation}
where $a$ is a non-negative constant and $a< \min_{i\in \mathds{N}}\frac{2\nu_i\xi_i^2}{N-3}$ for $N>3$, the uniqueness of the Nash equilibrium is ensured \cite{MaHuSpanos}.

Based on this given model, a Nash seeking strategy will be proposed in this section for the players to search for the unique Nash equilibrium (by assuming that $a< \min_{i\in \mathds{N}}\frac{2\nu_i\xi_i^2}{N-3}$ for $N>3$ in the rest of the paper).

\subsection{Nash Equilibrium Seeking for Energy Consumption Game of HVAC Systems}
\begin{Lemma}\label{lemm_11}
The energy consumption game is a potential game with a potential function being
\begin{equation}
Q(\mathbf{l})=\sum_{i=1}^{N}\nu_i\xi_i^2(l_i-\hat{l}_i)^2+\sum_{i=1}^{N}a(\sum_{j=1,j\neq i}^{N}l_j)l_i+\sum_{i=1}^N(al_i^{2}+p_0l_i).
\end{equation}
\end{Lemma}
\begin{Proof}
Noticing that
\begin{equation}
\frac{\partial Q(\mathbf{l})}{\partial l_i}=\frac{\partial C_i(\mathbf{l})}{\partial l_i}, \forall i\in \mathds{N},
\end{equation}
the conclusion can be derived by using the definition of potential game.
\end{Proof}

Based on the primal-dual dynamics in \cite{DurrCDC12}, the Nash seeking strategy for player $i$ is designed as
\begin{subequations}\label{upd_law_uni_lag}
\begin{align}
&\dot{D}_i=-D_i-\sum_{j \in \mathcal{N}_i}(D_{i}-D_{j})-\sum_{j \in \mathcal{N}_i}(\kappa_{i}-\kappa_{j})+Nl_i\\
&\dot{\kappa}_{i}=\sum_{j \in \mathcal{N}_i}(D_{i}-D_{j})\\
&\dot{l}_i=-\bar{k}_i(\frac{\partial V_i}{\partial l_i}+P(D_i)+al_i-\eta_{i1}+\eta_{i2})\label{upd_law_uni_1}\\
&\dot{\eta}_{i1}=\bar{m}_{i1}\eta_{i1}(l_i^{min}-l_i)\\
&\dot{\eta}_{i2}=\bar{m}_{i2}\eta_{i2}(l_i-l_i^{max}),i\in \mathds{N}
\end{align}
\end{subequations}
where $\bar{m}_{ij}=\delta m_{ij}$ for all $i \in \mathds{N}, j \in \{1,2\}$, $m_{ij}$ are fixed positive parameters and $\eta_{ij}(0)>0,i\in \mathds{N}, j\in \{1,2\}$.

By introducing the orthonormal matrix $U$ as in Section \ref{Nas_ex}, the system in \eqref{upd_law_uni_lag} can be rewritten as
\begin{subequations}\label{sa_law_ch}
\begin{align}
    \left[
    \begin{array}{c}
    \dot{\mathbf{D}} \\
    \dot{\widetilde{\mathbf{\kappa}}}
    \end{array}
    \right]&=\left[
    \begin{array}{cc}
    -I-L & -L \widetilde{U} \\
    \widetilde{U}^T L & \mathbf{0} \\
    \end{array}
    \right]
    \left[
    \begin{array}{c}
    \mathbf{D} \\
    \widetilde{\mathbf{\kappa}} \\
    \end{array}
    \right]+\left[
    \begin{array}{c}
    N\mathbf{l} \\
    \mathbf{0} \\
    \end{array}
    \right]\label{upd_law_ch_1} \\
    \dot{l}_i&=-\bar{k}_i(\frac{\partial V_i}{\partial l_i}+P(D_i)+al_i-\eta_{i1}+\eta_{i2})\\
 \dot{\eta}_{i1}&=\bar{m}_{i1}\eta_{i1}(l_i^{min}-l_i)\\
 \dot{\eta}_{i2}&=\bar{m}_{i2}\eta_{i2}(l_i-l_i^{max}),i\in \mathds{N}
\end{align}
\end{subequations}
and
\begin{equation}
\dot{\kappa}_\mu=0.
\end{equation}

\begin{Theorem}\label{asy_re}
Suppose that Assumption \ref{Ass_model} is satisfied.
Then, there exists $\beta \in \mathcal{KL}$ such that for each pair of
strictly positive real number $(\Delta ,v),$ there exists $\delta(\Delta ,v)>0,$  such that
\begin{equation}\label{kl_fun}
||\mathbf{\chi}(t)|| \leq \beta (|| \mathbf{\chi}(0)||,\delta t)+v,
\end{equation}%
for all $t\geq 0,$ $|| \mathbf{\chi}(0)|| \leq \Delta$ under \eqref{sa_law_ch} given that $\eta_{ij}(0)>0,i\in \mathds{N}, j\in \{1,2\}$. In \eqref{kl_fun}, $\mathbf{\chi}(t)=[(\mathbf{l}(t)-\mathbf{l}^*)^T,(\mathbf{D}(t)-\mathbf{1}\sum_{i=1}^Nl_i^*)^T, (\widetilde{\mathbf{\kappa}}(t)-\widetilde{\mathbf{\kappa}}^e(\mathbf{l}^*))^T,(\mathbf{\eta}(t)-\mathbf{\eta}^*)^T]^T$, and $\mathbf{\eta}$, $\mathbf{\eta}^*$ are defined in the subsequent proof.
\end{Theorem}
\begin{Proof}
Following the proof of Theorem \ref{res_1} by using singular perturbation analysis, the reduced-system in $\tau$-time scale is given by
\begin{equation}\label{redu_uni_ee}
\begin{aligned}
&\frac{dl_i}{d\tau}=-k_i(2\nu_i\xi_i^2(l_i-\hat{l}_i)+a\sum_{j=1}^{N}l_j+p_0+al_i-\eta_{i1}+\eta_{i2})\\
&\frac{d{\eta}_{i1}}{d\tau}=m_{i1}\eta_{i1}(l_i^{min}-l_i)\\
&\frac{d{\eta}_{i2}}{d\tau}=m_{i2}\eta_{i2}(l_i-l_i^{max}),i\in \mathds{N}.
\end{aligned}
\end{equation}

According to Lemma \ref{lemm_11}, the energy consumption game is a potential game. Hence, the Nash seeking can be achieved by solving
\begin{equation}\label{pot_min}
\begin{aligned}
\min_\mathbf{l} &\ \ Q(\mathbf{l}) \\
\text{subject to  } &l_i^{min}\leq l_i \leq l_i^{max}, i\in \mathds{N}.
\end{aligned}
\end{equation}

In the following, we show that \eqref{redu_uni_ee} can be used to solve the problem in \eqref{pot_min}.

Define the Lagrangian function as $L(\mathbf{l},\mathbf{\eta})=Q(\mathbf{l})+\sum_{i=1}^{N}(\eta_{i1}(l_i^{min}-l_i)+\eta_{i2}(l_i-l_i^{max}))$
where $\mathbf{\eta}=[\eta_{11},\eta_{12},\eta_{21},\eta_{22},\cdots,\eta_{i1},\eta_{i2},\cdots,\eta_{N1},\eta_{N2}]^T\in R^{2N}_{+}.$
The dual problem for the minimization problem in \eqref{pot_min} can be formulated as
\begin{equation}
\max_{\mathbf{\eta}\geq 0} \min_{\mathbf{l}} \ \ L(\mathbf{l},\mathbf{\eta}).
\end{equation}
The Hessian matrix of $Q(\mathbf{l})$ is
\begin{equation*}
H=\left[
  \begin{array}{cccc}
    2\nu_1\xi_1^2+2a & a & \cdots & a \\
    a & 2\nu_2\xi_2^2+2a & \cdots & a \\
    \vdots &  & \ddots & \vdots \\
    a & a & \cdots & 2\nu_N\xi_N^2+2a \\
  \end{array}
\right].
\end{equation*}
Since $a< \min_{i\in \mathds{N}}\frac{2\nu_i\xi_i^2}{N-3}$ for $N>3$, $H$ is positive definite by the Gershgorin Circle Theorem \cite{Horn85}. Hence, $Q(\mathbf{l})$ is strictly convex in $\mathbf{l}$ as its Hessian matrix is positive definite. Noting that the inequality constraints are linear, the problem in \eqref{pot_min} has strong duality \cite{Boyd04}. Hence, $\mathbf{l}^*$ is the optimal solution to the problem in \eqref{pot_min} if and only if there exists $\mathbf{\eta}^*\in R^{2N}_{+}$ such that $(\mathbf{l}^*, \mathbf{\eta}^*)$ is the saddle point of $L(\mathbf{l},\mathbf{\eta})$ by the saddle point theorem \cite{Boyd04}.

By defining the Lyapunov candidate function as \cite{DurrCDC12}
\begin{equation}
\begin{aligned}
V_L=&\frac{1}{2}(\mathbf{l}-\mathbf{l}^*)^T\mathbf{k}^{-1}(\mathbf{l}-\mathbf{l}^*)+\\
&\sum_{i=1}^{N}\sum_{j=1}^{2}\frac{1}{m_{ij}}(\eta_{ij}-\eta_{ij}^*-\eta_{ij}^*log(\eta_{ij})+\eta_{ij}^*log(\eta_{ij}^*)),
\end{aligned}
\end{equation}
where $(\mathbf{l}^*,\mathbf{\eta}^*)$ is the saddle point of $L(\mathbf{l},\mathbf{\eta})$, it can be shown that the saddle point of $L(\mathbf{l},\mathbf{\eta})$ is globally asymptotically stable under \eqref{redu_uni_ee} by Corollary 2 of \cite{DurrCDC12} given that $\eta_{ij}(0)>0, i\in \mathds{N}, j\in \{1,2\}$.

Hence, the strategy in \eqref{redu_uni_ee} enables $\mathbf{l}$ to converge to the Nash equilibrium of the potential game asymptotically.

Combining this result with the exponential stability of the boundary-layer system (see Theorem \ref{res_1} for boundary-layer analysis), the result can be derived by using Lemma 1 in \cite{TanAT06} (see also \cite{Teel} for more details).
\end{Proof}

\subsection{Nash Equilibrium Seeking for Energy Consumption Game of HVAC Systems with A Unique Inner Nash Equilibrium }
In the following, energy consumption game with a unique inner Nash equilibrium is considered \footnote{To make it clear, in this paper, we say that the Nash equilibrium is an inner Nash equilibrium if the Nash equilibrium satisfies $l_i^{min}<l_i^*<l_i^{max},\forall i\in \mathds{N}$, i.e., the Nash equilibrium is achieved at $\frac{\partial C_i}{\partial l_i}=0,\forall i\in \mathds{N}$.}. If the constraints do not affect the value of the Nash equilibrium, the Nash seeking strategy can be designed as
\begin{subequations}\label{upd_law_uni}
\begin{align}
&\dot{D}_i=-D_i-\sum_{j \in \mathcal{N}_i}(D_{i}-D_{j})-\sum_{j \in \mathcal{N}_i}(\kappa_{i}-\kappa_{j})+Nl_i\\
&\dot{\kappa}_{i}=\sum_{j \in \mathcal{N}_i}(D_{i}-D_{j})\\
&\dot{l}_i=-\bar{k}_i(\frac{\partial V_i}{\partial l_i}+P(D_i)+al_i).\label{upd_law_uni_1}
\end{align}
\end{subequations}

By introducing the orthonormal matrix $U$ as in Section \ref{Nas_ex}, the system in \eqref{upd_law_uni} can be rewritten as

\begin{subequations}\label{upd_law_uni_abc}
\begin{align}
    \left[
    \begin{array}{c}
    \dot{\mathbf{D}} \\
    \dot{\widetilde{\mathbf{\kappa}}}
    \end{array}
    \right]&=\left[
    \begin{array}{cc}
    -I-L & -L \widetilde{U} \\
    \widetilde{U}^T L & \mathbf{0} \\
    \end{array}
    \right]
    \left[
    \begin{array}{c}
    \mathbf{D} \\
    \widetilde{\mathbf{\kappa}} \\
    \end{array}
    \right]+\left[
    \begin{array}{c}
    N\mathbf{l} \\
    \mathbf{0} \\
    \end{array}
    \right] \\
    \dot{l}_i&=-\bar{k}_i(\frac{\partial V_i}{\partial l_i}+P(D_i)+al_i),i\in \mathds{N}
\end{align}
\end{subequations}
and
\begin{equation}
\dot{\kappa}_\mu=0.
\end{equation}

\begin{Theorem}\label{res_2}
Suppose that Assumption \ref{Ass_model} is satisfied. Then, there exists a positive constant $\delta^*$ such that for every $0<\delta<\delta^*,$
$(\mathbf{l}(t),\mathbf{D}(t),\widetilde{\mathbf{\kappa}}(t))$ converges exponentially to $(\mathbf{l}^*,\mathbf{1}\sum_{i=1}^{N}l_i^*,\widetilde{\mathbf{\kappa}}^e(\mathbf{l}^*))$
under \eqref{upd_law_uni_abc}.
\end{Theorem}
\begin{Proof}
Following the proof of Theorem \ref{res_1} by using singular perturbation analysis, the reduced-system at $\tau$-time scale is given by

\begin{equation}\label{redu_uni_e_1}
\begin{aligned}
\frac{dl_i}{d\tau}&=-k_i(2\nu_i\xi_i^2(l_i-\hat{l}_i)+a\sum_{j=1}^{N}l_j+p_0+al_i),i\in \mathds{N}.
\end{aligned}
\end{equation}
From \eqref{redu_uni_e_1}, it can be derived that
\begin{equation}
\dot{\widetilde{\mathbf{l}}}=-\mathbf{k}H\widetilde{\mathbf{l}},
\end{equation}
where $\widetilde{\mathbf{l}}(t)=\mathbf{l}-\mathbf{l}^*$.

Hence, it can be shown that in $\tau$-time scale
\begin{equation}
\begin{aligned}
||\widetilde{\mathbf{l}}(\tau)||&\leq \sqrt{\frac{\max_{i\in \mathbb{N}}\{k_i\}}{\min_{i\in \mathbb{N}}\{k_i\}}}e^{-\min_{i\in \mathbb{N}}\{k_i\}\lambda_{min}\{H\}\tau}||\widetilde{\mathbf{l}}(0)||,
\end{aligned}
\end{equation}
by defining the Lyapunov candidate function as $V_L=\frac{1}{2}\widetilde{\mathbf{l}}^T\mathbf{k}^{-1}\widetilde{\mathbf{l}}$ \cite{Frihauf12}.

Combining this result with the exponential stability of the boundary-layer system (see Theorem \ref{res_1} for boundary-layer analysis), the conclusion can be derived
by Theorem 11.4 in \cite{KHAIL02} (see the proof of Theorem \ref{res_1} for more details) .
\end{Proof}
\begin{Remark}
When the constraints do not affect the value of the Nash equilibrium, the updating strategy in \eqref{upd_law_uni} is a special case of the one in \eqref{upd_law}. It can be seen that the result in Theorem \ref{res_2} is stronger than the result in Theorem \ref{res_1} under the given HVAC model.
\end{Remark}

\subsection{Energy Consumption Game of HVAC Systems with Stubborn Players}

In this section, a special case where some players commit to the coordination process while keeping a constant energy consumption is considered. Without loss of generality, we suppose that player $i$ is a stubborn player and updates its action according to
\begin{subequations}\label{upd_law_stu_5}
\begin{align}
&\dot{D}_i=-D_i-\sum_{j \in \mathcal{N}_i}(D_{i}-D_{j})-\sum_{j \in \mathcal{N}_i}(\kappa_{i}-\kappa_{j})+Nl_i^s,\\
&\dot{\kappa}_{i}=\sum_{j \in \mathcal{N}_i}(D_{i}-D_{j}),
\end{align}
\end{subequations}
where $l_i^s$ is the constant energy consumption of player $i$. Furthermore, all the rational players adopt \eqref{upd_law_uni} if the constraints do not affect the value of all the players' best response strategies, else, all the rational players adopt \eqref{upd_law_uni_lag}.

By introducing the orthonormal matrix $U$ as in Section \ref{Nas_ex}, then

\begin{equation}\label{red}
    \left[
    \begin{array}{c}
    \dot{\mathbf{D}} \\
    \dot{\widetilde{\mathbf{\kappa}}}
    \end{array}
    \right]=\left[
    \begin{array}{cc}
    -I-L & -L \widetilde{U} \\
    \widetilde{U}^T L & \mathbf{0} \\
    \end{array}
    \right]
    \left[
    \begin{array}{c}
    \mathbf{D} \\
    \widetilde{\mathbf{\kappa}} \\
    \end{array}
    \right]+\left[
    \begin{array}{c}
    N\mathbf{l} \\
    \mathbf{0} \\
    \end{array}
    \right],
\end{equation}
in which the $i$th component of $\mathbf{l}$ is fixed to be $l_i^s,$
and
\begin{equation}
\dot{\kappa}_\mu=0.
\end{equation}

Furthermore,
\begin{equation}\label{red_2}
\begin{aligned}
&\dot{l}_j=-\bar{k}_j(\frac{\partial V_j}{\partial l_j}+P(D_j)+al_j-\eta_{j1}+\eta_{j2})\\
&\dot{\eta}_{j1}=\bar{m}_{j1}\eta_{j1}(l_j^{min}-l_j)\\
&\dot{\eta}_{j2}=\bar{m}_{j2}\eta_{j2}(l_j-l_j^{max}),j\in \mathds{N}, j\neq i
\end{aligned}
\end{equation}
if all the rational players adopt \eqref{upd_law_uni_lag}, and
\begin{equation}\label{red_3}
\dot{l}_j=-\bar{k}_j(\frac{\partial V_j}{\partial l_j}+P(D_j)+al_j), j\in \mathds{N}, j\neq i
\end{equation}
if all the rational players adopt \eqref{upd_law_uni}.
\begin{Corollary}
Suppose that Assumption \ref{Ass_model} is satisfied.
Then, there exists $\beta \in \mathcal{KL}$ such that for each pair of
strictly positive real number $(\Delta ,v),$ there exists $\delta(\Delta ,v)>0,$  such that
\begin{equation}\label{kl_fun_1}
||\mathbf{\chi}^{br}(t)|| \leq \beta (|| \mathbf{\chi}^{br}(0)||,\delta t)+v,
\end{equation}%
for all $t\geq 0,$ $|| \mathbf{\chi}^{br}(0)|| \leq \Delta$ under \eqref{red} and \eqref{red_2} given that $\eta_{kj}(0)>0, k\in \mathds{N}, k\neq i, j\in \{1,2\}$. In \eqref{kl_fun_1}, $\mathbf{\chi}^{br}(t)=[\check{\mathbf{l}}_{-i}(t)^T, (\mathbf{D}(t)-\mathbf{1}(\sum_{j=1,j\neq i}^N l_j^{br}+l_i^s))^T,(\widetilde{\mathbf{\kappa}}(t)-\widetilde{\mathbf{\kappa}}^e(\mathbf{l}_{-i}^{br},l_i^s))^T,(\mathbf{\eta}_{-i}(t)-\mathbf{\eta}_{-i}^{br})^T]^T$, $\check{\mathbf{l}}_{-i}(t)=[\check{l}_1(t),\cdots,\check{l}_{i-1}(t),\check{l}_{i+1}(t),\cdots,\check{l}_{N}(t)]^T,$ $\check{l}_j(t)=l_j(t)-l_j^{br},j\in \mathds{N}, j\neq i$ and $l_j^{br}$ denotes the best response strategy of player $j$, $\mathbf{l}_{-i}^{br}$ is the concatenated vector of $l_j^{br},i\in \mathds{N}, j\neq i,$ $\mathbf{\eta}_{-i}$, $\mathbf{\eta}_{-i}^{br}$ are defined in the subsequent proof.

Furthermore, if the constraints $l_j^{min}\leq l_j \leq l_j^{max},j\in \mathds{N},j\neq i$ do not affect the values of the best response strategies, then, there exists a positive constant $\delta^*$ such that for every $0<\delta<\delta^*,$ $(\check{\mathbf{l}}_{-i}(t), \mathbf{D}(t)-\mathbf{1}(\sum_{j=1,j\neq i}^N l_j^{br}+l_i^s),\widetilde{\mathbf{\kappa}}(t)-\widetilde{\mathbf{\kappa}}^e(\mathbf{l}_{-i}^{br},l_i^s))$ converges exponentially to zero under \eqref{red} and \eqref{red_3}.
%
%
\end{Corollary}
\begin{Proof} Following the proof of Theorem \ref{res_1} by using singular perturbation analysis to get the reduced-system for both cases.
 The first part of the Corollary can be derived by noticing that the following is satisfied,
\begin{equation}
\frac{\partial Q^{'}(\mathbf{l}_{-i})}{\partial l_j}=\frac{\partial C_j}{\partial l_j},
\end{equation}
for all $j\neq i$, where
\begin{equation}
\begin{aligned}
Q^{'}(\mathbf{l}_{-i})=&\sum_{j=1,j\neq i}^{N}\nu_j\xi_j^2(l_j-\hat{l}_j)^2
+\sum_{j=1,j\neq i}^{N}a(\sum_{k=1,k\neq i,k\neq j}^{N}l_k)l_j\\
&+\sum_{j=1,j\neq i}^{N}(al_j^{2}+(p_0+al_i^{s})l_j).
\end{aligned}
\end{equation}
Define
\begin{equation}
\begin{aligned}
L^{'}({\mathbf{l}_{-i},\mathbf{\eta}_{-i}})&=Q^{'}(\mathbf{l}_{-i})
+\sum_{j=1,j\neq i}^N\eta_{j1}(l_j^{min}-l_j)\\
&+\sum_{j=1,j\neq i}^N\eta_{j2}(l_j-l_j^{max}),
\end{aligned}
\end{equation}
where $\mathbf{\eta}_{-i}\in R^{2(N-1)}_{+}$ denotes the concatenated vector of $\eta_{jk},j\in \mathds{N}$ and $j\neq i$, $k\in \{1,2\}.$

Then, the dual problem is
\begin{equation*}
\max_{\mathbf{\eta}_{-i} \geq 0} \min_{\mathbf{l}_{-i}} L^{'}({\mathbf{l}_{-i},\mathbf{\eta}_{-i}}).
\end{equation*}
Noticing that $\mathbf{l}_{-i}^{br}$ is the best response strategies of the rational players if and only if there exists $\mathbf{\eta}_{-i}^{br}\in R^{2(N-1)}_{+}$ such that $(\mathbf{l}_{-i}^{br},\mathbf{\eta}_{-i}^{br})$ is the saddle point of $L^{'}({\mathbf{l}_{-i},\mathbf{\eta}_{-i}})$, the rest of the proof follows that of Theorem \ref{asy_re}.

If the constraints $l_j^{min}\leq l_j \leq l_j^{max}, j\in \mathds{N}, j\neq i$ do not affect the value of the best response strategies and all the rational players adopt \eqref{upd_law_uni}, then, the reduced-system at the quasi-steady state is

\begin{equation}\label{stab_eq2}
\begin{aligned}
\frac{dl_j}{d\tau}&=-k_j \frac{\partial C_j(\mathbf{l}_{-i})}{\partial l_j}\\
&=-k_j[2\nu_j\xi_j^2(l_j-\hat{l}_j)+a\sum_{j=1,j\neq i}^{N}l_j+p_0+al_i^s+al_j]\\
&=-k_j[2\nu_j\xi_j^2\check{l}_j+a\sum_{j=1,j\neq i}^{N}\check{l}_j+a\check{l}_j], i\in \mathds{N}.
\end{aligned}
\end{equation}
Writing (\ref{stab_eq2}) in the concatenated form gives
\begin{equation}
\frac{d\check{\mathbf{l}}_{-i}}{d\tau}=-\mathbf{k}H_1\check{\mathbf{l}}_{-i},
\end{equation}
where $H_1\in R^{N-1}\times R^{N-1}$ is defined as
\begin{equation*}
H_1=\small{\left[
  \begin{array}{cccccc}
    h_1 & a & \cdots  &  &a\\
     &  \ddots&  &  &\\
     &  & h_{i-1}   & &\\
     &  & &h_{i+1}  &\vdots\\
    \vdots &  &  &  \ddots &  & \\
    a &  &  \cdots  & & h_N\\
  \end{array}
\right]},
\end{equation*}
with $h_j=2\nu_j\xi_j^2+2a.$
Since $a<\min_{i\in \mathds{N}}\frac{2\nu_i\xi_i^2}{N-3}$ for $N>3$, it can be derived that the matrix $H_1$ is symmetric and strictly diagonally dominant with all the diagonal elements positive. Hence, the matrix $-\mathbf{k}H_1$ is Hurwitz by the Gershgorin Circle Theorem \cite{Horn85}.

By similar analysis in Theorem \ref{res_2}, the conclusion can be derived.
\end{Proof}

\begin{Remark}
In this Corollary, only one player (i.e., player $i$) is supposed to be stubborn. However, this is not restrictive as similar results can be derived if multiple stubborn players exist.
\end{Remark}
\begin{Remark}
In the proposed Nash seeking strategy, the players only communicate with its neighbors on $D_i$ and $\kappa_i$. They do not communicate on their own energy consumption $l_i$. Hence, the proposed Nash seeking strategy does not lead to privacy concern for the users. The work in \cite{Frihauf12} provided an extremum seeking method to seek for the Nash equilibrium in non-cooperative games. However, the method in \cite{Frihauf12} can't be directly implemented for Nash seeking in the energy consumption game if the aggregate energy consumption is not directly available to the players.
\end{Remark}

\section{Numerical Examples}\label{num_e}
\subsection{Simulation Setup}\label{simu_set}
\begin{figure}
\begin{center}
\scalebox{0.35}{\includegraphics[21,326][398,565]{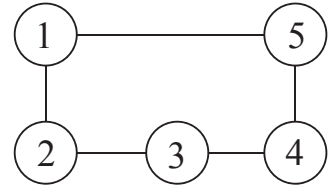}}
\caption{Communication graph for the electricity users}\label{commu_graph}
\end{center}
\end{figure}

In this section, we consider a network of $5$ commercial/industrial users that are equipped with HVAC systems. The electricity users communicate with each other via an undirected and connected graph as shown in Fig. \ref{commu_graph}.
The cost function for electricity user $i$ is
\begin{equation}
C_{i}(l_{i},\bar{l})=\nu_i\xi_i^2(l_i-\hat{l}_i)^2+P(\bar{l})l_{i},
\end{equation}
where the pricing function $P(\bar{l})=a\sum_{i=1}^{N} l_i+p_0$ \cite{MaHuSpanos}. Without loss of generality, suppose that $\nu_i\xi_i^2$ for all $i \in \mathds{N}$ are normalized to $1$ in the simulation. For $i\in\{1,2,3,4,5\}$, $l_i^{min}=0.8\hat{l}_i$ and $l_i^{max}=1.2\hat{l}_i$ except that in Section \ref{HVAC_SE}, $l_1^{min}$ and $l_1^{max}$ are set separately. The parameters are listed in Table \ref{tab_1}\footnote{MU stands for Monetary Unit.}.
\begin{table}
\caption{Parameters in the simulation.}
\label{tab_1}
\begin{center}
\begin{tabular}{|c|c|c|c|c|c|}
  \hline
   & user 1 & user 2 & user 3 & user 4 & user 5 \\
   \hline
  $\hat{l}_i$(kWh) & 50 & 55 & 60 & 65 & 70 \\
  \hline
  $l_i^{max}$(kWh) & 60 & 66 & 72 & 78 & 84 \\
  \hline
  $l_i^{min}$(kWh) & 40 & 44 & 46 & 52 & 56 \\
  \hline
  $a$ &\multicolumn{5}{c|}{0.04}  \\
  \hline
  $p_0$ (MU/kWh) &\multicolumn{5}{c|}{5} \\
  \hline
\end{tabular}
\end{center}
\ \ \ \ \ \ \\
\ \ \ \ \ \ \\
\end{table}

\subsection{Energy Consumption Control of HVAC Systems}\label{HVAC_SE}
In this section, we suppose that $l_1^{min}=45$ and $l_1^{max}=55$.  By direct computation, it can be derived that the Nash equilibrium is $\mathbf{l}^*=(45, 46.4, 51.3, 56.2, 61.1)$(kWh). The equilibrium aggregate is $\bar{l}^*=259.9$(kWh). Hence, the Nash equilibrium is not an inner Nash equilibrium.

\begin{figure}
\begin{center}
\scalebox{0.18}{\includegraphics[29,60][1008,541]{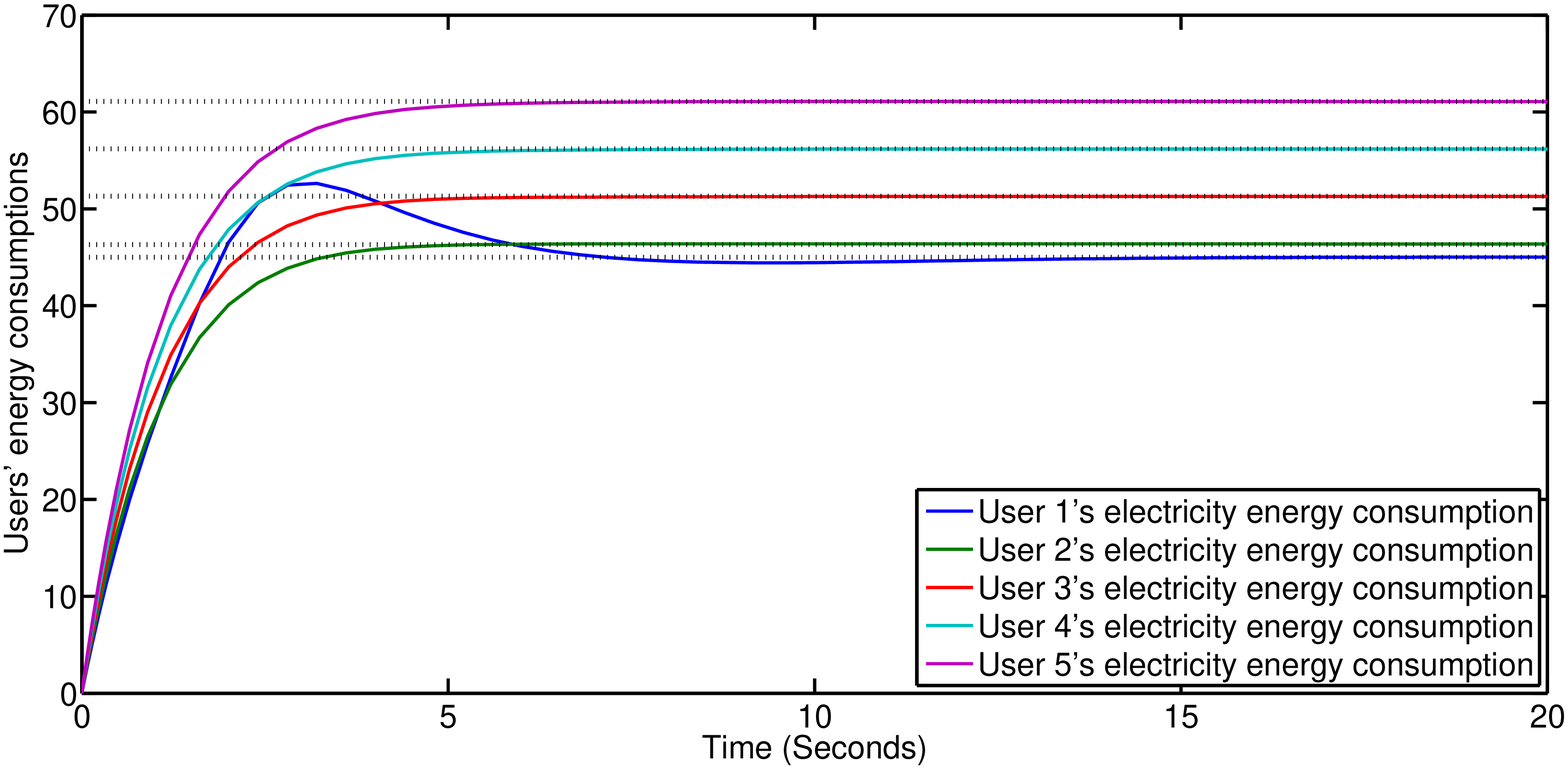}}
\caption{The users' energy consumptions produced by the proposed seeking strategy in \eqref{upd_law_uni_lag}.}
\label{load_cons}
\end{center}
\end{figure}

\begin{figure}[thpb]
\begin{center}
\ \ \ \\
\ \ \\
\scalebox{0.18}{\includegraphics[16,49][1027,561]{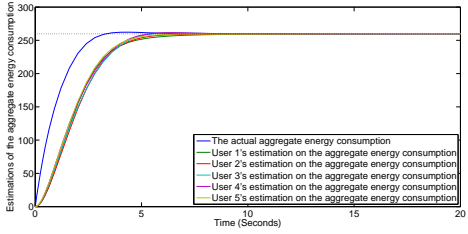}}
\caption{The actual aggregate energy consumption and the users' estimations on the aggregate energy consumption produced by the proposed seeking strategy in \eqref{upd_law_uni_lag}.}
\label{agge_lo_cons}
\end{center}
\end{figure}

The simulation results by using the seeking strategy in \eqref{upd_law_uni_lag} are shown in Figs. \ref{load_cons}-\ref{agge_lo_cons}.

Fig. \ref{load_cons} shows the users' electricity energy consumptions produced by the proposed seeking strategy in \eqref{upd_law_uni_lag} and Fig. \ref{agge_lo_cons} indicates that the users' estimations on the aggregate energy consumptions converge to the actual aggregate energy consumption.

From the simulation results, it can be seen that the energy consumptions produced by the proposed method converge to the Nash equilibrium of the energy consumption game.

\subsection{Energy Consumption Control of HVAC Systems with A Unique inner Nash equilibrium}
\begin{figure}[hp]
\begin{center}
\ \ \\
\ \ \\
\ \ \\
\scalebox{0.18}{\includegraphics[1,50][1018,593]{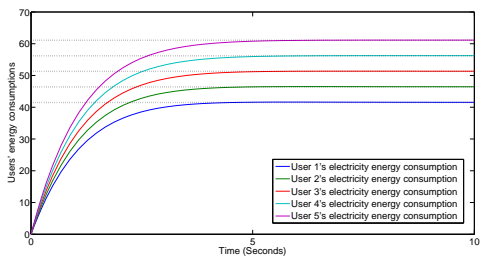}}
\caption{The users' energy consumptions produced by the proposed seeking strategy in \eqref{upd_law_uni}.}
\label{load_j}
\end{center}
\end{figure}

\begin{figure}[hpb]
\begin{center}
\ \ \\
\scalebox{0.168}{\includegraphics[8,44][1026,612]{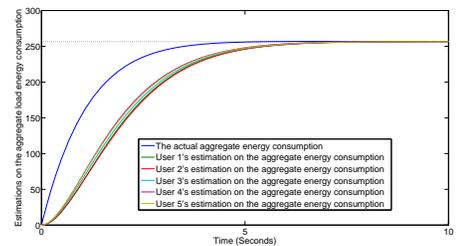}}
\caption{The actual aggregate energy consumption and the users' estimations on the aggregate energy consumptions produced by the seeking strategy in \eqref{upd_law_uni}.}
\label{agge_lo_j}
\end{center}
\end{figure}

In this section, we consider the energy consumption game under the setting in Section \ref{simu_set}. By direct calculation, it can be derived  that the Nash equilibrium is $\mathbf{l}^*=(41.5,46.4,51.3,56.2,61.1)$(kWh). The equilibrium aggregate is $\bar{l}^*=256.7$(kWh). The Nash equilibrium is an inner Nash equilibrium and the seeking strategy in \eqref{upd_law_uni} is used in the simulation. The simulation results produced by the seeking strategy in \eqref{upd_law_uni} are shown in Figs. \ref{load_j}-\ref{agge_lo_j}.

From the simulation results, it can be seen that the users' energy consumptions converge to the unique Nash equilibrium.

\subsection{Energy Consumption Control with Stubborn Players}
\begin{figure}[thpb]
\begin{center}
\scalebox{0.18}{\includegraphics[32,50][1020,563]{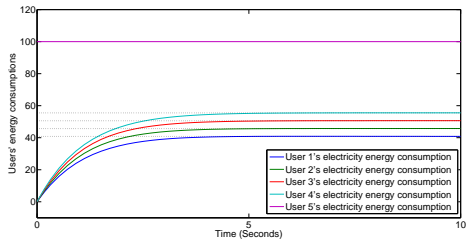}}
\caption{The users' energy consumptions produced by the proposed method (i.e., the rational players adopt \eqref{upd_law_uni} and the stubborn player adopts \eqref{upd_law_stu_5}).}
\label{st_loads}
\end{center}
\end{figure}

\begin{figure}[hpb]
\begin{center}
\ \ \\
\ \ \\
\scalebox{0.18}{\includegraphics[35,77][1023,556]{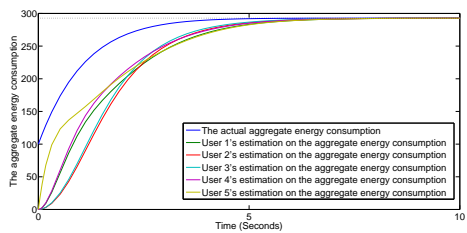}}
\caption{The actual aggregate energy consumption and the users' estimations on the aggregate energy consumption produced by the proposed method (i.e., the rational players adopt \eqref{upd_law_uni} and the stubborn player adopts \eqref{upd_law_stu_5}).}
\label{st_agg}
\end{center}
\end{figure}

In this section, we suppose that player $5$ is a stubborn player that commits to a constant energy consumption $l_5^s=100$(kWh). Then, player $1$-$4$'s best response strategies are $40.8$(kWh), $45.7$(kWh), $50.6$(kWh), and $55.5$(kWh), respectively. The aggregate energy consumption is $292.7$(kWh). In the simulation, the rational players adopt \eqref{upd_law_uni} to update their actions. The stubborn player uses the seeking strategy in \eqref{upd_law_stu_5}. The simulation results produced by the proposed method are shown in Figs. \ref{st_loads}-\ref{st_agg}.

It can be seen that with the presence of the stubborn player, all the other players' actions converge to the best response strategies with respect to the stubborn action.

\section{Conclusions and Future Work}\label{con_a}
This paper considers energy consumption control among a network of electricity users. The problem is solved by using an aggregative game on an undirected and connected graph. To estimate the aggregate energy consumption, which is supposed to be unknown to the players during the Nash seeking process, an average consensus protocol is employed. The convergence property is analytically studied via using singular perturbation and Lyapunov stability analysis. A general energy consumption game where multiple Nash equilibria may exist is firstly considered. A Nash seeking strategy based on consensus is proposed to enable the users to search for the Nash equilibrium. Energy consumption control of HVAC systems with linear pricing functions is then studied. Convergence results are provided. Furthermore, stubborn players are investigated and it is shown that the rational players' actions converge to the best response strategies.

For future directions, the following aspects would be considered:
\begin{enumerate}
  \item The design of incentive provoking mechanisms. As Nash solution is usually not efficient from the system-level perspective, socially optimal solution might be preferred if coordination is allowed. Incentive provoking mechanisms can be designed to motivate the electricity users to coordinate such that system efficiency can be improved \cite{YETCST}.
  \item Analysis of the energy consumption game with the existence of cheaters. This includes the detection of cheaters, the design of penalty (e.g., \cite{MaTII15}) or reward algorithms to prevent cheating behaviors, etc.
  \item Nash seeking for energy consumption game under various communication conditions.
\end{enumerate}

\end{document}